\documentclass[singlecolumn,notitlepage,amsmath,amssymb,aps,prl,groupedaddress,floatfix,showpacs,superscriptaddress]{revtex4-1}

\usepackage{graphicx}
\usepackage{dcolumn}
\usepackage{bm}
\usepackage{hyperref}
\hypersetup{colorlinks=true,urlcolor=blue,citecolor=blue}
\usepackage[usenames, dvipsnames]{color} 
\usepackage{braket} 
\usepackage{soul}
\usepackage{tabularx}
\usepackage[caption=false]{subfig}

\begin{document}
\title{Utilizing machine learning to improve the precision of fluorescence imaging of cavity-generated spin squeezed states} 
\author{Benjamin K. Malia}
\affiliation{Department of Physics, Stanford University, Stanford, California 94305, USA}
\author{Yunfan Wu}
\affiliation{Department of Applied Physics, Stanford University, Stanford, California 94305, USA}
\author{Juli\'{a}n Mart\'{i}nez-Rinc\'{o}n}
\affiliation{Department of Physics, Stanford University, Stanford, California 94305, USA}
\author{Mark A. Kasevich}
\email[]{kasevich@stanford.edu}
\affiliation{Department of Physics, Stanford University, Stanford, California 94305, USA}
\affiliation{Department of Applied Physics, Stanford University, Stanford, California 94305, USA}
\date{\today}

\begin{abstract}
We present a supervised learning model to calibrate the photon collection rate during the fluorescence imaging of cold atoms. The linear regression model finds the collection rate at each location on the sensor such that the atomic population difference equals that of a highly precise optical cavity measurement. This 192 variable regression results in a measurement variance 27\% smaller than our previous single variable regression calibration. The measurement variance is now in agreement with the theoretical limit due to other known noise sources. This model efficiently trains in less than a minute on a standard personal computer's CPU, and requires less than 10 minutes of data collection. Furthermore, the model is applicable across a large changes in population difference and across data collected on different days. 
\end{abstract}

\maketitle

\section{Introduction}
Machine learning (ML) is becoming an increasingly important tool for analysis of scientific data due to its ability to handle large data sets with many dependant variables. In supervised learning, a regression model can be trained on previously collected data to find relationships between an experimental result and potentially dozens of parameters. A successful ML model will be able to accurately predict future data from these many parameters.

A variety of ML algorithms are able to process high resolution images. In experiments involving fluorescence imaging, ML has recently been applied to denoise images~\cite{Wang2021}, classify objects in images~\cite{Sagar2020}, classify spectral signatures~\cite{Ju2019}, and quantify fluorescent decay lifetimes~\cite{Mannam2020,Smith2019}. In cold atom experiments, fluorescence imaging is used for atom number measurement to determine the populations of different quantum states. A variety of challenges, including scattered light, sensor read noise, and inhomogeneous photon collection, increase the difficulty of obtaining accurate measurements~\cite{Rocco_2014}. The need for accurate imaging is vital in the context of quantum metrology. In these experiments, imaging imperfections fundamentally limit the efficacies of quantum protocols. For example, Qu et al.~\cite{Qu2020} used principle component analysis to correct for background light in order to fully characterize a quantum state.
\section{Spin Squeezing Measurements}
Atomic clocks and sensors operate through measurement of population differences between quantum states. For two spin states $\ket{\uparrow}$ and $\ket{\downarrow}$, the collective spin is defined as $J_z = (N_{\ket{\uparrow}}-N_{\ket{\downarrow}})/2$ where $N = N_{\ket{\uparrow}} + N_{\ket{\downarrow}}$ is the total number of atoms in the sensor. The changes in $J_z$ are measured through two consecutive measurements, one before and one after the sensor has detected a field. We define the population difference as $J_z^{(1,2)}= J_z^{(2)}-J_z^{(1)}$.

In the absence of environmental fields, a single measurement of $J_z^{(1,2)}$ should be close to, but not exactly equal to, zero due to the Heisenberg uncertainty principal. In a typical sensor, without spin-squeezing, the quantum projection noise (QPN) limits the variance of the measurement difference to $\big(\Delta J_z^{(1,2)}\big)^2=N/4$ ~\cite{Hosten_2016,Cox_2016,Schleier-Smith_2010b}. If the quantum state is squeezed with a quantum non-demolition (QND) measurement, as in our experiment, then the variance in population difference is reduced such that $\Delta J_z^{(1,2)}$ surpasses the QPN. While $J_z^{(1)}$ is a QND measurement, $J_z^{(2)}$ can be either another QND measurement or a fluorescence measurement. Squeezing allows for increased sensor precision when increasing atom number or interrogation time is not possible.

Here we employ ML to identify and correct for a class of imperfects associated with non-uniform spatial imaging profiles. Potential sources which may contribute to non-uniform spatial imaging are the angle of light entering the objective lens ~\cite{Catrysse2002}, inhomogeneous transparency of vacuum chamber walls, and an inhomogeneous imaging beam power.  As we show below, these imperfections, inferred to be ~10\% across a mm-scale spatially extended cloud, led to a 33\% increase in the noise variance in the work of Ref.~\cite{Malia2020}.

To evaluate the performance of the system as a sensor, we define this difference in terms of angular difference $\theta = J_z^{(1,2)}/C(N/2)$ on the multiparticle Bloch sphere, where $C$ is the contrast in a Ramsey sequence. In the case of the data in ~\cite{Malia2020}, which is the subject of our analysis in this paper, $C=0.92$ and the QPN is defined as $\Delta\theta = 1/\sqrt{N}$. As an example for a sensor with $N=390\,000$, the QPN is $\Delta\theta = 1.6~\text{mrad}$. Eliminating extra noise sources is crucial to taking advantage of the noise reduction from spin squeezing.

The relevant details of our experiment~\cite{Malia2020} are summarized here. ${}^{87}\text{Rb}$ atoms are cooled and trapped in an optical lattice. The QND measurement is first performed by detecting a cavity-QED dispersive readout phase using an optical light puslse. The lattice is then turned off and the atoms free fall. To perform population spectroscopy at the end of the free fall time, an on-resonance laser pulse imparts a momentum kick to the atoms in the $\ket{\uparrow}$ state. After the two states spatially separate, a retroreflected laser induces fluorescence in both states. The light is collected with an objective lens with a numerical aperture of 0.25 and is sent to a CMOS camera. Each state takes up approximately half of the sensor region. $J_z^{(2)}$ is determined by the difference in total counts between states $\ket{\uparrow}$ and $\ket{\downarrow}$.

As a result of the mechanisms discussed earlier, the average number of counts per atom in each state are several percent different. Due to shot-to-shot fluctuations in the atoms' positions, the spatial inhomogeneity in the scattering rate directly leads to fluctuations in the inferred atomic populations. This additional noise is large enough that, uncorrected, results in $\Delta J_z$ larger than the QPN.

\section{Machine Learning Model}
\begin{figure}[h]
\centering
\includegraphics[width=0.7\linewidth]{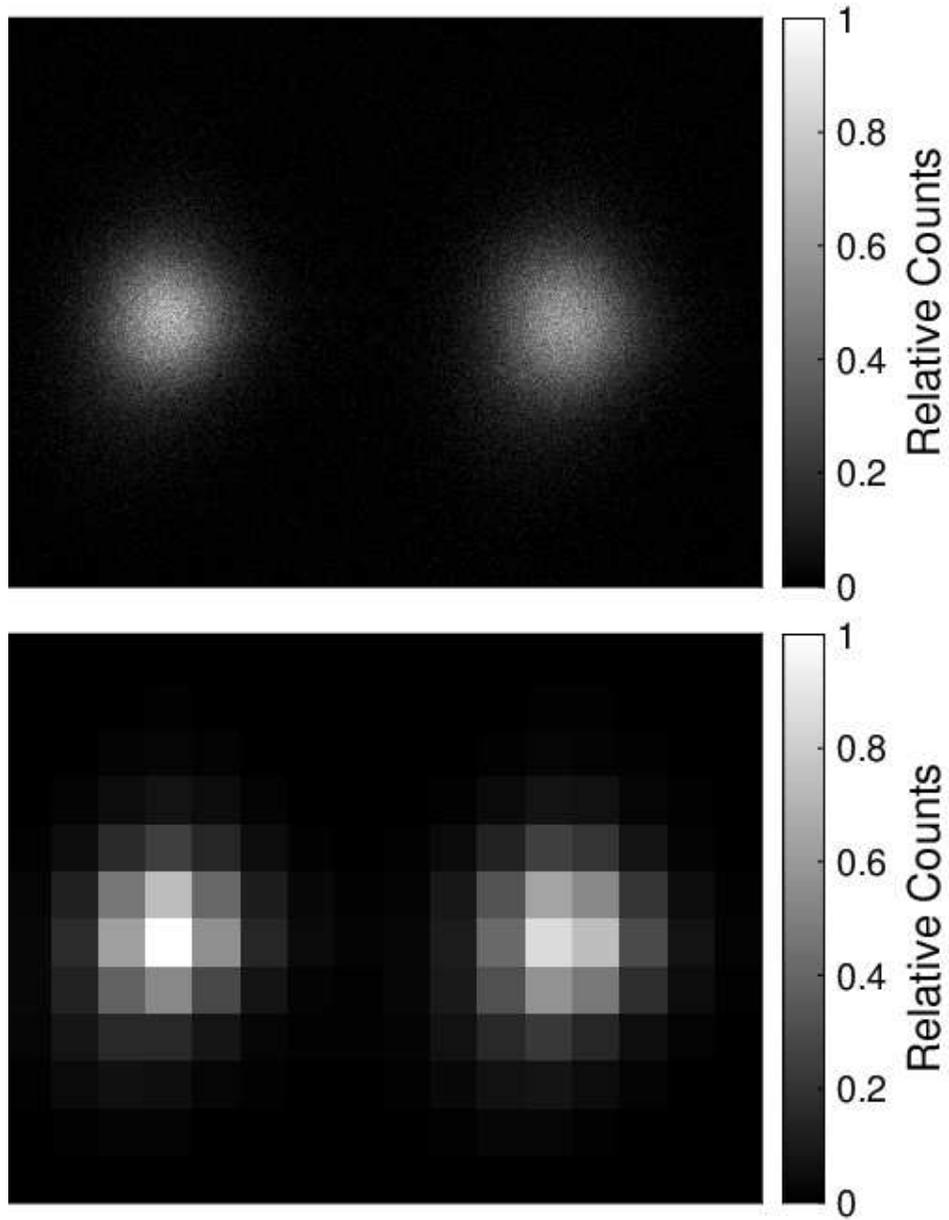}
\caption{Relative counts for each pixel. Full resolution image with $3.1\times 10^6$ pixels (upper) and counts binned into 192 superpixels (lower). The left half of the image ($\ket{\downarrow}$) is subtracted from the right half ($\ket{\uparrow}$) to obtain the population difference.}
\label{fig:counts}
\end{figure}

The supervised learning model estimates $J_z^{(2)}$ by first dividing the fluorescence image into 192 ``superpixels" (Figure~\ref{fig:counts}). These superpixels are simply $128\times128$ pixel squares grouped into blocks. Our new regression model now has $n=192$ variables to fit to. Each superpixel is then assigned a weight $\beta_j$ which is multiplied by its total counts $c_j$ to estimate the correct atom number for that pixel. The estimated $\tilde{J}_z^{(2),(i)}$ are determined by subtracting the estimated atoms in the left half of the image from the ones in the right half:
\begin{equation}
    \tilde{J}_z^{(2),(i)} = \beta_0 + \frac{1}{2}\sum_{j=1}^{n/2} c^{(i)}_j\beta_j- \frac{1}{2}\sum_{j=n/2+1}^n c^{(i)}_j\beta_j
\end{equation}
where $i$ index the images and $\beta_0$ is a bias term. We can also determine the total atom number $N$ from these $\beta_j$:
\begin{equation}
    \tilde{N}^{(i)} = \sum_{j=1}^{n} c^{(i)}_j\beta_j
\end{equation}

The inhomogeneity must at least vary on length scales smaller than the atom cloud to make a significant impact on the precision of the measurement. The maximum spatial frequency accessible to the model is determined by the resolution of the images. We therefore chose a superpixel resolution large enough to capture most of the spatial variation, but small enough to allow for fast training on small sample sizes. As described below, we enforce a penalty term which puts an upper bound on the spatial frequency of the model's result.

Consecutive low noise cavity measurements show that detection limit of the squeezed state is $\Delta \theta=310~\mu \text{rad}$, which is lower than the 690~$\mu$rad theoretical limit of the fluorescence detection (see \cite{Malia2020} supplement for details on these noise sources). We therefore take $J_z^{(1)}$ as the target for $J_z^{(2)}$ when training our regression model. 

We calculate the optimal $\beta_j$ by minimizing the cost function, $G$, with the Broyden–Fletcher–Goldfarb–Shanno algorithm~\cite{Broyden1970} over $m$ samples. Specifically, the \emph{fminunc} function in MATLAB is supplied with both $G$ and its gradient. $G$ is defined as a weighted least squares with regularization~\cite{Neumaier1998}:
\begin{equation}
    G = \frac{1}{2m_c}\sum_{i=1}^{m} w^{(i)} \big(\tilde{J}_z^{(2),(i)}-\tilde{J}_z^{(1),(i)}\big)^2 + \frac{\lambda}{2m_c}P_{NN}.
    \label{eq:cost}
\end{equation}
Here, the weights $w^{(i)}$ are defined by the Heaviside step function $H$ and the magnitudes of $J^{(1),(i)}_z$:
\begin{equation}
    w^{(i)} = H\big(|J_z^{(1),(i)}| - J_z^\text{cutoff}\big),
\end{equation}
$m_c$ is the total number of samples with non-zero weight. $\tilde{J}_z^{(1),(i)}$ is the frequency-corrected cavity measurement (\cite{Malia2020} supplement, Equation S8)
\begin{equation}
    \tilde{J}_z^{(1),(i)} = J^{(1),(i)}_z + \frac{\delta^{(i)}}{2\Delta}\tilde{N}^{(i)}  
\end{equation}
where $\delta^{i}$ is the difference in frequency between the probe laser and the target frequency (detuned $\Delta$ from the $\ket{\uparrow}\rightarrow\ket{e}$ transition). The nearest neighbor penalty is
\begin{equation}
    P_{NN} = \sum_{<a,b>} (\beta_b-\beta_a)^2
\end{equation}
where $<a,b>$ represents the set of all unique nearest neighbor pairs of superpixels and $\lambda$ is the regularization hyperparameter. The $(\Delta\theta)^2$ predicted by this model is the variance of $\tilde{\theta} = (\tilde{J}_z^{(2)}-\tilde{J}_z^{(1)})/(C \tilde{N}/2)$.

\section{Results and Discussion}
\begin{figure}[h]
\centering
\includegraphics[width=0.7\linewidth]{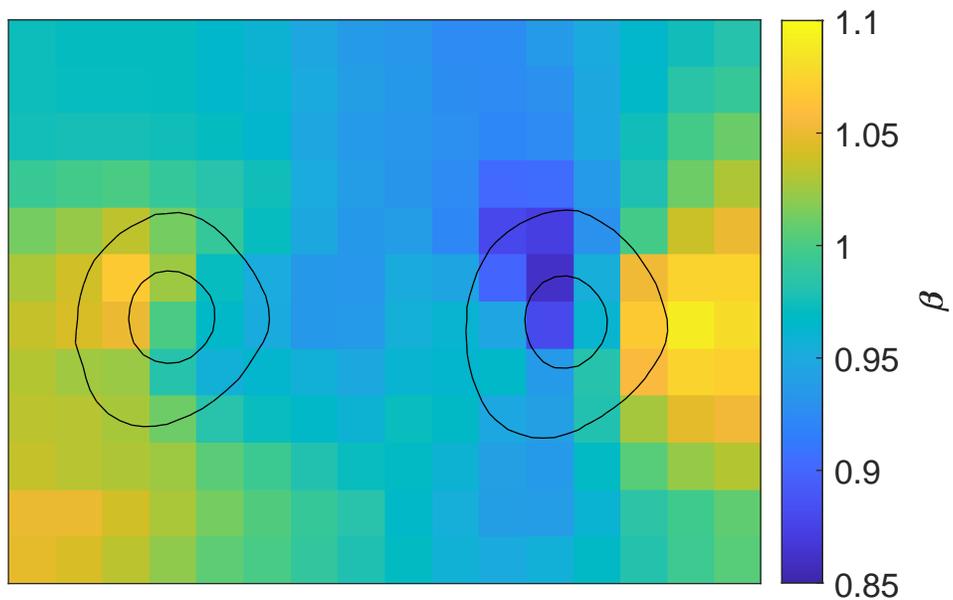}
\caption{Relative values for the weights $\beta_j$ for each superpixel under the parameters $J_z^\text{cutoff}=200$ and $\lambda = 20$. Overlay: inner curves enclose 68\% of atoms and outer curves enclose 95\% of atoms in a sample image (upper Figure~\ref{fig:counts}).}
\label{fig:map}
\end{figure}

\begin{figure}[h]
\centering
\includegraphics[width=0.7\linewidth]{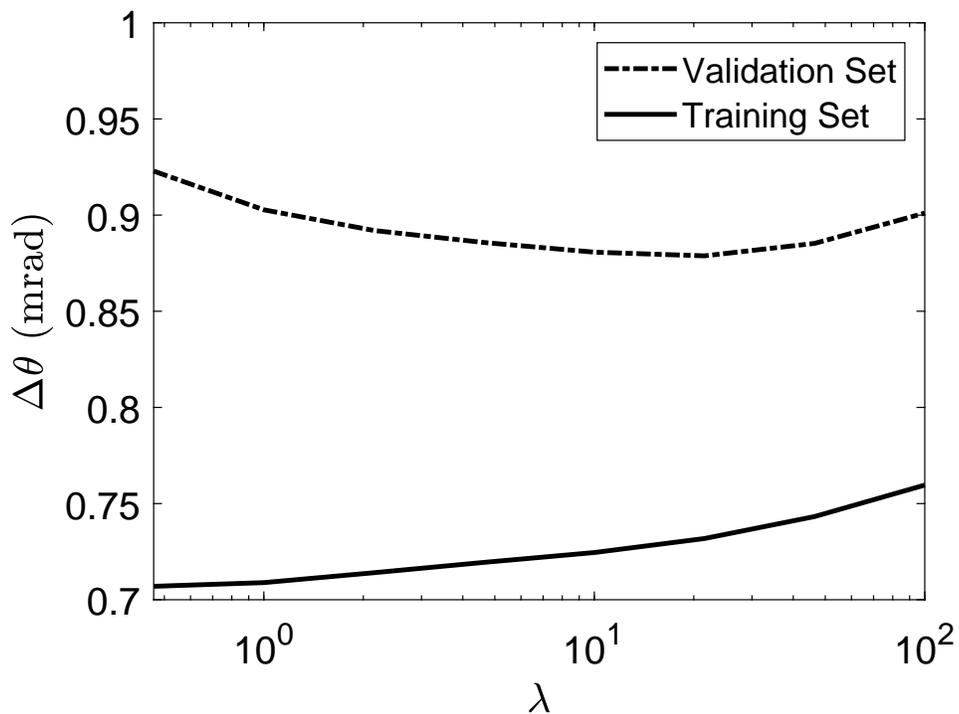}
\caption{The evaluation metric $\Delta\theta$ is used to choose the weight $\lambda$ of the nearest neighbor penalty. The optimal regularization is $\lambda=20$ when $J_z^\text{cutoff}=200$.}
\label{fig:lambda}
\end{figure}

\begin{figure}[h]
\centering
\includegraphics[width=0.7\linewidth]{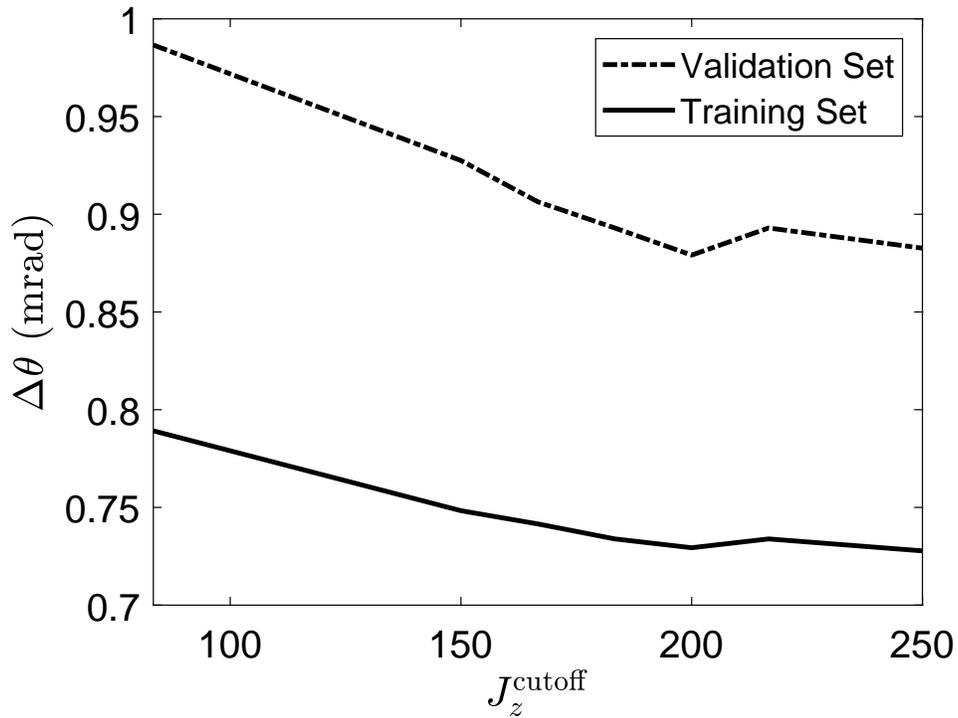}
\caption{The evaluation metric $\Delta\theta$ is used to choose the hyperparameter $J_z^\text{cutoff}$. The optimal value is $J_z^\text{cutoff}=200$ when $\lambda = 20$.}
\label{fig:cutoff}
\end{figure}

We train the model on a set of 500 images taken from the data collected for Ref.~\cite{Malia2020}. 120 images remain after $J_z^\text{cutoff}$ is applied. The hyperparameters, $J_z^\text{cutoff}$ and $\lambda$, are chosen to give the smallest $\Delta\theta$ in a separate validation set of 50 images. Figures~\ref{fig:lambda} and~\ref{fig:cutoff} show the results of training with a range of values for each hyperparameter. The optimal values are $J_z^\text{cutoff}=200$ and $\lambda = 20$. Note that samples with $|J_z^{(1)}|$ this large are typically excluded when the system is used as a sensor as those $J_z^{(1)}$ closest to zero are most accurate. Therefore, the validation set will have larger $\Delta\theta$ than that of the final result. The $\beta_j$ determined with these parameters are shown in Figure~\ref{fig:map}. $\beta_0 < 10^{-3}$ and does not significantly contribute to $\tilde{J}_z^{(2)}$.

\begin{figure}[h]
\centering
\includegraphics[width=0.7\linewidth]{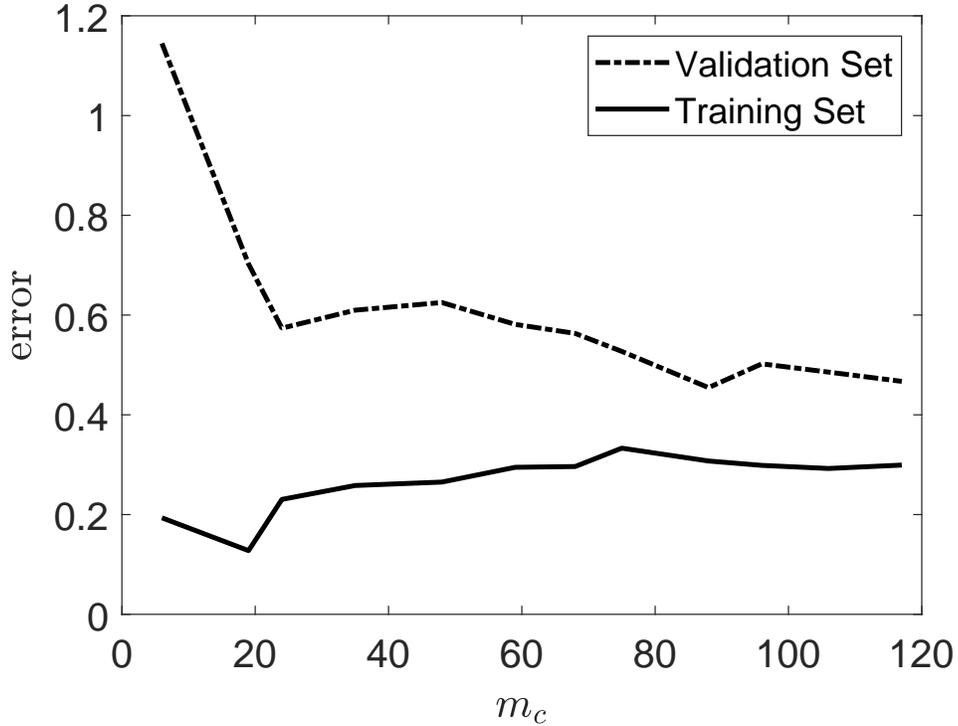}
\caption{Mean least squares error (first term of the cost function $G$) vs number of samples with nonzero weight ($m_c$). Optimal values of $J_z^\text{cutoff}=200$ and $\lambda = 20$ are used. Final difference between sets is 0.17.}
\label{fig:gap}
\end{figure}

Figure~\ref{fig:gap} shows the learning curves as a function of nonzero weight sample size ($m_c$). Relatively few data points are needed for the model to minimize the difference of mean least squares error (the first term in Equation~\ref{eq:cost}) between the training and validation sets. The maximum number of points used ($m_c = 117$) corresponds to approximately 8 minutes of data collection and 1 minute of training. Increasing the model complexity by increasing the number of superpixels does not significantly change the gap between the training and validation set errors at maximum $m_c$. It also does not change the qualitative structure of the $\beta$ distribution. On the other hand, a model with less complexity fails to minimize the gap.

At the cost of sample size, using the Heaviside function, $H$, to set the weights prevents the model from optimizing on values near zero. This eliminates a $\tilde{J}^{(1)}_z$ dependent trend in the residual $\tilde{J}_z^{(2)}-\tilde{J}_z^{(1)}$ and an unreasonably small $\tilde{N}^{(i)}$. When weights are included, the mean of predicted atom numbers $\tilde{N}^{(i)}$ are within 2\% of the previously calibrated mean value. The nearest neighbor penalty acts similarly to ridge regularization~\cite{Neumaier1998} in that it prevents the model from overfitting and keeps a relatively similar contribution from each $\beta_j$. The optimal $\lambda$ reveals the characteristic spatial frequency of the inhomogeneity. 

Despite the model fitting only on samples with $|J_z^{(1),(i)}|$ larger than the cutoff, it extrapolates successfully onto those with values smaller than the cutoff. Applying the model to the remaining data (samples with $|J_z^{(1),(i)}|<J_z^\text{cutoff}$) results in an angular resolution of $\Delta\theta = 691\pm60~\mu\text{rad}$. This variance is 27\% smaller than our previous reported result (see Table ~\ref{tab:testValues}). It is also consistent with our previously predicted limit given the known noise sources. In terms of metrologically useful squeezing, this is $7.2\pm0.3$ dB below the QPN limit.

This collection efficiency map is independent of the mean $J_z$ as well as the mean position of the atoms and the total atom number. It is therefore versatile and can be applied to data sets taken on different days with different parameters. For example, with measurements where the atomic state is rotated along the polar angle of the Bloch sphere by 25 mrad and the total atom number is reduced by 40\%, the model now predicts $\Delta\theta = 880\pm60~\mu\text{rad}$ ($7.3\pm0.3$ dB below the QPN), a 27\% reduction in variance upon the previous method (see Table ~\ref{tab:testValues}).

The collection inhomogeneity across the cloud can also lead to a degradation in the correlation between a single image and its corresponding QND measurement~\cite{Hu2015}. In this work, the estimated increase in variance due to this noise source is 0.5\%~\cite{Wu2020}, which is negligible. The ML model demonstrated here primarily reduces the variance from shot-to-shot fluctuations in the atom cloud distributions coupled with the detection inhomogeneity.

In our previously reported results we applied a single weight, determined by the mean position of the atoms, to all of the counts from the $\ket{\uparrow}$ state. Although this method provided significant improvement (Table~\ref{tab:testValues}), the atoms are spread out over most of the image so applying the same same factor to all counts is an over-simplification. ML provides the capability of increasing the complexity of our model to accurately calibrate the measurement.
\section{Conclusion}
In this work, we have shown that supervised learning is capable of characterizing inhomogeneous fluorescence measurements when prior low noise measurements are available for training the model. This model is robust to changes in spatial arrangement of the atomic populations and correctly predicts the total atom number. This method could be extended to absorption imaging of cold atoms. Absorption imaging, especially high intensity imaging, requires careful calibration to accurately count atoms~\cite{Gross2012,Hueck2017}. A ML approach could circumvent traditional calibration techniques to convert pixel-by-pixel intensity changes to atom numbers in spin squeezing experiments~\cite{Huang2020}.

Other ML techniques have the potential to further improve our result. In Ness et al.~\cite{Ness2020}, deep learning is used to predict the background noise based on the signal surrounding the atoms. A similar approach may allow our procedure to forgo subtracting the second background image. This would reduce the variance introduced by the background by up to half of its current value.

\section{Acknowledgments}
\begin{acknowledgments}
This work is supported by the Department of Energy (DE-SC0019174-0001) and the Vannevar Bush Faculty Fellowship.
\end{acknowledgments}

\clearpage
\section{Figures}

\clearpage
\section{Tables}
\begingroup
\squeezetable
\begin{table}[h]
\begin{tabularx}{0.5\linewidth}{c @{\hskip 0.35cm}c @{\hskip 0.35cm}c@{\hskip 0.35cm} c}
\hline
 \hline
 \begin{tabular}{c}mean $\tilde{\theta}^{(2)}$ \\(mrad)\end{tabular} & \begin{tabular}{c}no correction\\(mrad) \end{tabular}&\begin{tabular}{c} mean position\\ correction (mrad)\end{tabular} & \begin{tabular}{c}supervised \\learning (mrad)\end{tabular}\\
\hline
 0 & $1.86\pm0.09$ &$0.81\pm0.06$ &$0.69\pm0.06$  \\ 

 25 &$2.23\pm0.15$ &$1.03\pm0.06$&$0.88\pm0.06$  \\ 
 \hline
 \hline
\end{tabularx}
\caption{Angular resolution $\Delta\theta$ of the test sets for different analysis methods. The set where the fluorescence measurement has a mean value of $\tilde{\theta}^{(2)}=\tilde{J}_z^{(2)}/C(\tilde{N}/2)=0$ contains $N=390\,000$ (QPN limit is $\Delta\theta=1.60~\text{mrad}$). The set with $\theta^{(2)}=25~\text{mrad}$ contains $N=240\,000$ (QPN limit is $\Delta\theta=2.04~\text{mrad}$).}
\label{tab:testValues}
\end{table}
\endgroup

\clearpage
\section{Pseudo code for training model}

\noindent \textcolor{red}{BEGIN SCRIPT}\\
\newline
\noindent\textcolor{green}{\% Import experimental results}\\
\noindent \textcolor{blue}{SET} \textcolor{magenta}{CavityJz} to the Jz values from the QND measurement of the sample dataset\\
\noindent \textcolor{blue}{SET} \textcolor{magenta}{PixelValuesReal} to the counts in the images corresponding to each \textcolor{magenta}{CavityJz}\\
\noindent \textcolor{blue}{SET} \textcolor{magenta}{FrequencyShift} to the difference in probe frequency between the measured value and the target value\\
\noindent \textcolor{blue}{DIVIDE}  \textcolor{magenta}{FrequencyShift} by half of the hyperfine transition frequency\\
\newline
\noindent\textcolor{green}{\% Create superpixels}\\
\noindent \textcolor{blue}{FOR} each image\\
	\indent \indent \textcolor{blue}{FOR} each 128x128 section of real pixels\\
		\indent \indent \indent \indent \textcolor{blue}{SET} \textcolor{magenta}{PixelValues} to sum of \textcolor{magenta}{PixelValuesReal} inside the section\\
    \indent\indent \textcolor{blue}{END FOR}\\
\noindent \textcolor{blue}{END FOR}\\
\newline
\noindent\textcolor{green}{\% Segment experimental data}\\
\noindent\textcolor{blue}{SORT} \textcolor{magenta}{CavityJz}, \textcolor{magenta}{PixelValues}, and \textcolor{magenta}{FrequencyShift} into Training, Validaiton, and Test datasets\\
\newline
\noindent\textcolor{green}{\% Initialize and minimize beta}\\
\noindent \textcolor{blue}{SET} \textcolor{magenta}{InitialBetaBias} to 0\\
\noindent \textcolor{blue}{SET} all \textcolor{magenta}{InitialBetaTrain} to 1\\
\newline
\noindent \textcolor{blue}{CALL} \textcolor{cyan}{fminunc} with \textcolor{cyan}{EvaluateCost}, \textcolor{magenta}{InitialBetaBias}, \textcolor{magenta}{InitialBetaTrain}, \textcolor{magenta}{PixelValuesTrain}, \textcolor{magenta}{CavityJzTrain}, \textcolor{magenta}{FrequencyShiftTrain}\\
    \indent\indent \textcolor{blue}{RETURNING} \textcolor{magenta}{FinalBiasBeta}, \textcolor{magenta}{FinalBeta}, \textcolor{magenta}{FinalCost}\\
\newline
\textcolor{green}{\% Find the model results for the Validation and Test sets}\\
\noindent \textcolor{blue}{CALL} \textcolor{cyan}{EvaluateModel} with \textcolor{magenta}{FinalBiasBeta}, \textcolor{magenta}{FinalBeta}, \textcolor{magenta}{PixelValuesValidation}\\
	\indent \indent \textcolor{blue}{RETURNING} \textcolor{magenta}{JzValidation},  \textcolor{magenta}{NValidation}\\
\noindent \textcolor{blue}{SET} \textcolor{magenta}{TargetJzValidation} to \textcolor{magenta}{CavityJzValidation} plus half the product of \textcolor{magenta}{FrequencyShiftValidation} and \textcolor{magenta}{NValidation}\\
\noindent \textcolor{blue}{SET} \textcolor{magenta}{VarianceJzValidation} to the variance of the difference between \textcolor{magenta}{JzValidation} and \textcolor{magenta}{TargetJzValidation}\\
\newline
\noindent \textcolor{blue}{CALL} \textcolor{cyan}{EvaluateModel} with \textcolor{magenta}{FinalBiasBeta}, \textcolor{magenta}{FinalBeta}, \textcolor{magenta}{PixelValuesTest}\\
	\indent \indent \textcolor{blue}{RETURNING} \textcolor{magenta}{JzTest}, \textcolor{magenta}{NTest}\\
\noindent \textcolor{blue}{SET} \textcolor{magenta}{TargetJzTest} to \textcolor{magenta}{CavityJzTest} plus half the product of \textcolor{magenta}{FrequencyShiftTest} and \textcolor{magenta}{NTest}\\
\noindent \textcolor{blue}{SET} \textcolor{magenta}{VarianceJzTest} to the variance of the difference between \textcolor{magenta}{JzTest} and \textcolor{magenta}{TargetJzTest}\\
\textcolor{red}{END SCRIPT}\\
\newline
\newline
\textcolor{red}{FUNCTION} \textcolor{cyan}{fminunc}\\
\noindent \textcolor{blue}{PASS IN} cost function and its arguments (\textcolor{cyan}{EvaluateCost}, \textcolor{magenta}{BetaBias}, \textcolor{magenta}{BetaTrain}, \textcolor{magenta}{PixelValues}, \textcolor{magenta}{CavityJz}, \textcolor{magenta}{FrequencyShift})\\
\newline
\noindent\textcolor{green}{\% This is a built in MATLAB function based on the BFGS algorithm.}\\
\noindent\textcolor{green}{\% The options used are:}\\
\noindent\textcolor{green}{\% TolX = 1E-7}\\
\noindent\textcolor{green}{\% TolFun = 1E-7}\\
\noindent\textcolor{green}{\% MaxIter 1E5}\\
\noindent\textcolor{green}{\% MaxFunction = 5E5}\\
\newline
\noindent\textcolor{blue}{WHILE} model error is larger than tolerances\\
	\indent\indent \textcolor{blue}{CALL} \textcolor{cyan}{EvaluateCost} with \textcolor{magenta}{BetaBias}, \textcolor{magenta}{BetaTrain}, \textcolor{magenta}{PixelValues}, \textcolor{magenta}{CavityJz}, \textcolor{magenta}{FrequencyShift}\\
		\indent\indent\indent\indent \textcolor{blue}{RETURNING} \textcolor{magenta}{Cost}, \textcolor{magenta}{GradientCost}, \textcolor{magenta}{GradientCostBias}\\
	\indent\indent \textcolor{blue}{COMPUTE} BFGS algorithm\\
	\indent\indent\textcolor{blue}{UPDATE} \textcolor{magenta}{BetaBias} and \textcolor{magenta}{BetaTrain}\\
\textcolor{blue}{END WHILE}\\
\newline
\noindent \textcolor{blue}{PASS OUT} \textcolor{magenta}{BetaBias}, \textcolor{magenta}{BetaTrain}, \textcolor{magenta}{Cost}\\
\textcolor{red}{END FUNCTION}\\
\newline
\noindent\textcolor{red}{FUNCTION}  \textcolor{cyan}{EvaluateCost}\\
\noindent \textcolor{blue}{PASS IN} \textcolor{magenta}{BetaBias}, \textcolor{magenta}{BetaTrain}, \textcolor{magenta}{PixelValues}, \textcolor{magenta}{CavityJz}, \textcolor{magenta}{FrequencyShift}\\
\newline
\noindent\textcolor{green}{\% Initialize variables}\\
\noindent \textcolor{blue}{SET} all \textcolor{magenta}{JzWeights} to 0\\
\noindent \textcolor{blue}{SET} \textcolor{magenta}{NeigborPenalty} to 0\\
\noindent \textcolor{blue}{SET} all \textcolor{magenta}{NeighborGradients} to 0\\
\noindent \textcolor{blue}{SET} \textcolor{magenta}{JzNorm} to the mean of \textcolor{magenta}{CavityJz}\\
\noindent \textcolor{blue}{SET} \textcolor{magenta}{Cost} to 0\\
\noindent \textcolor{blue}{SET} all \textcolor{magenta}{GradientCost} to 0\\
\noindent \textcolor{blue}{SET} \textcolor{magenta}{SubsetSize} to 0\\
\newline
\noindent\textcolor{green}{\% Define hyperparameters}\\
\noindent \textcolor{blue}{SET} \textcolor{magenta}{NeighborWeight} to 20\\
\noindent \textcolor{blue}{SET} \textcolor{magenta}{JzCutoff} to 200\\
\newline
\noindent \textcolor{blue}{FOR} each sample\\
    \indent\indent\textcolor{green}{\% Find this iteration's Jz and N values for the Test set}\\
	\indent \indent \textcolor{blue}{CALL} \textcolor{cyan}{EvaluateModel} with \textcolor{magenta}{BetaBias}, \textcolor{magenta}{BetaTrain}, and \textcolor{magenta}{PixelValues} \\
		\indent\indent\indent\indent \textcolor{blue}{RETURNING} \textcolor{magenta}{TestJz}, \textcolor{magenta}{TestN}\\
    \newline
	\indent\indent\textcolor{green}{\% Apply frequency correction and find error}\\
	\indent\indent \textcolor{blue}{SET} \textcolor{magenta}{TargetJz} to \textcolor{magenta}{CavityJz} plus half the product of \textcolor{magenta}{FrequencyShift} and \textcolor{magenta}{TestN}\\
	\indent\indent \textcolor{blue}{SET} \textcolor{magenta}{JzError} to the difference of \textcolor{magenta}{TestJz} and \textcolor{magenta}{TargetJz}\\ 
	\indent \indent \textcolor{blue}{SET} \textcolor{magenta}{JzNorm} to the mean of \textcolor{magenta}{CavityJz}\\
	\indent \indent \textcolor{blue}{SET} \textcolor{magenta}{JzErrorNorm} to \textcolor{magenta}{JzError} to divided by \textcolor{magenta}{JzNorm}\\
    \newline
	\indent\indent\textcolor{blue}{IF} \textcolor{magenta}{TargetJz} is less then \textcolor{magenta}{JzCutoff}\\
		\indent\indent\indent\indent \textcolor{blue}{SET} \textcolor{magenta}{JzWeights} for this point to be 0\\
	\indent\indent\textcolor{blue}{ELSE} \\
	\indent\indent\indent\indent	\noindent \textcolor{blue}{SET} \textcolor{magenta}{JzWeights} for this point to be 1\\
		\indent\indent\indent\indent\textcolor{blue}{INCREMENT} \textcolor{magenta}{SubSetSize} by 1\\
	    \indent\indent\textcolor{blue}{END IF}\\

	\indent\indent\textcolor{green}{\% Find neighbor penalty and its gradient}\\
	\indent\indent \textcolor{blue}{FOR} each edge between neighboring pixels\\
	\indent\indent\indent\indent\textcolor{blue}{ADD} the square of the difference in \textcolor{magenta}{BetaTrain} (corresponding to the pixels touching this edge) to \textcolor{magenta}{NeigborPenalty}\\
	\indent\indent \textcolor{blue}{END FOR} \\
    \newline
	\indent\indent\noindent \textcolor{blue}{FOR} each pixel\\
	\indent\indent	\indent\indent\textcolor{blue}{ADD} the difference in \textcolor{magenta}{BetaTrain} to \textcolor{magenta}{NeighborGradients} for all edges touching this pixel\\
	\indent\indent\noindent \textcolor{blue}{END FOR}\\
    \newline
    \indent\indent\textcolor{green}{\% Calculate cost function and its gradient}\\
	\indent\indent \textcolor{blue}{ADD} the product of \textcolor{magenta}{JzWeights} the square of \textcolor{magenta}{JzError} to \textcolor{magenta}{Cost}\\
	\indent\indent\textcolor{blue}{SET} \textcolor{magenta}{GradientCostBias} to the product of \textcolor{magenta}{JzWeights}, and \textcolor{magenta}{JzError}\\
	\indent\indent \textcolor{blue}{FOR} each pixel corresponding the the left side of image\\
		\indent\indent\indent\indent\textcolor{blue}{SUBTRACT} the product of \textcolor{magenta}{JzWeights}, \textcolor{magenta}{JzError}, and \textcolor{magenta}{PixelValues(pixel)} from \textcolor{magenta}{GradientCost}\\
	\indent\indent \textcolor{blue}{END FOR}\\
	\indent\indent \textcolor{blue}{FOR} each pixel corresponding to the right side of image\\
		\indent\indent\indent\indent \textcolor{blue}{ADD} the product of \textcolor{magenta}{JzWeights}, \textcolor{magenta}{JzError}, and \textcolor{magenta}{PixelValues(pixel)} to \textcolor{magenta}{GradientCost}\\
	\indent\indent \textcolor{blue}{END FOR}\\
\noindent \textcolor{blue}{END FOR}\\
\newline
\noindent\textcolor{green}{\% add regularization amd normalize results}\\
\textcolor{blue}{ADD} the product of \textcolor{magenta}{NeigborPenalty} and \textcolor{magenta}{NeighborWeight} to \textcolor{magenta}{Cost}\\
\noindent \textcolor{blue}{DIVIDE} \textcolor{magenta}{Cost} by twice the \textcolor{magenta}{SubsetSize}\\
\newline
\noindent \textcolor{blue}{FOR} every sample\\
	\indent\indent \textcolor{blue}{FOR} every pixel\\
	\indent\indent	\indent\indent \textcolor{blue}{ADD} the product of twice \textcolor{magenta}{NeighborWeight} and \textcolor{magenta}{NeighborGradients} to \textcolor{magenta}{GradientCost}\\
	\indent\indent \textcolor{blue}{END FOR}\\
    \noindent \textcolor{blue}{END FOR}\\
\noindent\textcolor{blue}{DIVIDE} all \textcolor{magenta}{GradientCost} by twice the \textcolor{magenta}{SubsetSize}\\
\newline
\noindent \textcolor{blue}{PASS OUT} \textcolor{magenta}{Cost}, \textcolor{magenta}{GradientCost}, \textcolor{magenta}{GradientCostBias}\\
\textcolor{red}{END FUNCTION} \\
\newline
\textcolor{red}{FUNCTION} \textcolor{cyan}{EvaluateModel}\\
\noindent \textcolor{blue}{PASS IN} \textcolor{magenta}{BetaBias}, \textcolor{magenta}{BetaTrain}, \textcolor{magenta}{PixelValues}\\
\newline
\noindent\textcolor{green}{\% Initialize Variables}\\
\noindent \textcolor{blue}{SET} \textcolor{magenta}{Jz} to \textcolor{magenta}{BetaBias}\\
\noindent \textcolor{blue}{SET} \textcolor{magenta}{N} to 0\\
\newline
\noindent\textcolor{green}{\% Calculate N}\\
\noindent \textcolor{blue}{FOR} each pixel\\
\indent\indent\textcolor{blue}{ADD} product of \textcolor{magenta}{PixelValues(pixel)} and \textcolor{magenta}{BetaTrain(pixel)} to \textcolor{magenta}{N}\\
\noindent \textcolor{blue}{END FOR}\\
\newline
\noindent\textcolor{green}{\% Calculate Jz}\\
\noindent \textcolor{blue}{FOR} each pixel corresponding the the left side of image\\
\indent\indent\textcolor{blue}{SUBTRACT} product of \textcolor{magenta}{PixelValues(pixel)} and \textcolor{magenta}{BetaTrain(pixel)} from \textcolor{magenta}{Jz}\\
\noindent \textcolor{blue}{END FOR}\\
\noindent \textcolor{blue}{FOR} each pixel corresponding to the right side of image\\
\indent\indent\textcolor{blue}{ADD} product of \textcolor{magenta}{PixelValues(pixel)} and \textcolor{magenta}{BetaTrain(pixel)} to \textcolor{magenta}{Jz}\\
\noindent \textcolor{blue}{END FOR}\\
\indent\indent\textcolor{blue}{DIVIDE} \textcolor{magenta}{Jz} by 2\\
\newline
\noindent \textcolor{blue}{PASS OUT} \textcolor{magenta}{Jz}, \textcolor{magenta}{N}\\
\textcolor{red}{END FUNCTION} 
\end{document}